\def\fr#1#2{\hbox{${#1\over #2}$}}

\def\ni{\noindent}
\def\vs{\vskip.3cm}

\def\+{{(+)}}  \def\-{ {(-)} }   \def\0{ {(0)} }
\def\1{ {(1)} }  \def\2{ {(2)} }

\def\m{\mu}

\def\sq{Q\kern-6pt/}
\def\sQ{Q\kern-12pt\nearrow}
\documentclass[11pt,eqs]{article}

\usepackage{latexsym}
\usepackage{amssymb}
\def\be{\begin{equation}}             \def\ee{\end{equation}}
\def\ba{\begin{array}{rcl}}           \def\ea{\end{array}}
\def\beqa{\begin{eqnarray} }          \def\eeqa{\end{eqnarray} }
\def\beqalign{\begin{eqalign}}        \def\eeqalign{\end{eqalign}}
             
\def\bsubeq{\begin{subequations}}     \def\esubeq{\end{subequations}}
\def\bitem{\begin{itemize}}           \def\eitem{\end{itemize}}

\def\DJ{\leavevmode\setbox0=\hbox{D}\kern0pt
 \rlap{\kern.04em\raise.188\ht0\hbox{-}}D}
\def\dj{\leavevmode\setbox0=\hbox{d}\kern0pt
 \rlap{\kern.215em\raise.46\ht0\hbox{-}}d}

\newcommand{\bd}{\begin{displaymath}}
\newcommand{\ed}{\end{displaymath}}
\begin{document}

\title{ String theory landscape and cosmological constant
\thanks{Work supported in part by the Serbian Ministry of Science and
Technological Development.}}
\author{B. Nikoli\'c \thanks{e-mail address: bnikolic@ipb.ac.rs}\\{\it Institute of Physics Belgrade, University of Belgrade, Serbia}}
%\date{}
\maketitle

\begin{abstract}
In this paper we considered the bosonic string action in the presence of metric $G_{\mu\nu}$, Kalb-Ramond field $B_{\mu\nu}$ and dilaton field $\Phi$. The quantum conformal invariance is achieved if all three one-loop $\beta$-functions are zero. But, $\beta^\Phi$, $\beta$-function related to the dilaton field, can also be equal to some nonzero constant $c$, which is actually the central charge of Virasoro algebra. The anomaly contained in the Schwinger term of Virasoro algebra is canceled by adding Liouville term to the sigma model action and quantum conformal invariance is restored. 
The space-time action, which Euler-Lagrange equations of motion are quantum conformal invariance conditions with nonzero $\beta^{\Phi}$, is Einstein-Hilbert action with some matter and cosmological constant. The cosmological constant is obtained within landscape framework as linear function of the central charge of Virasoro algebra.

\end{abstract}
\vs

\ni {Keywords: cosmology constant, string landscape}  \par

\section{Introduction}

Einstein general theory of relativity is the geometric theory of
gravitation. It describes gravity as a geometric property of
space-time. In particular, the curvature of space-time is directly
related to the energy-momentum tensor of matter and radiation. The relation
is specified by the Einstein field equations, a system of partial
differential equations \cite{Land}.

In 1915, when Einstein finished his general theory of relativity,
the Universe was considered as a static one \cite{misner}. From
Einstein equations of gravitational field it follows that Universe
is a dynamic system. Facing with this inconsistency of the theory
and ruling philosophical beliefs that Universe is static, Einstein
changed his theory adding to the Einstein tensor,
$R_{\mu\nu}-\frac{1}{2}G_{\mu\nu}R$, term $\Lambda G_{\mu\nu}$,
where $\Lambda$ is so called cosmological constant \cite{sean}. Modified gravitational equations gives a static Universe as one of the solutions. But a few years
later, expansion of the Universe was discovered by Hubble and
confirmed that Universe was dynamic. It seemed that cosmological
constant was no longer necessary.  But in 1998 two independent observations of the distant type Ia supernova \cite{tipIa} proved that expansion of Universe is accelerated. Acceleration of Universe expansion is possible only if cosmological constant $\Lambda$ is positive, which means that vacuum density $\rho_{vac}$ is positive and pressure is negative. So, the cosmological constant came back in the story.                                

In the 60s of the previous century string theory occurred as a theory of strong interactions. Because of some unwanted features (tachyons, 26 dimensions), it was rejected as a theory of strong interactions. Shortly afterwards, about ten years later, string theory was again in the focus of physics but now as a theory-good candidate for unification of all known interactions. Later, fermions are incorporated in the theory and superstring theory was born. 

Bosonic string theory is consistent in 26, while superstring theory is consistent in 10 dimensional space-time. The consistency conditions can be considered as equations of motion for background fields and it is possible, in general, to write down the space-time action which produces these equations. Such action is called action of the effective theory (effective action) and it contains gravity. That is one of the reasons to consider superstring theory as good candidate for unification of all interactions. Also we can expect that such theory can give some interpretation of the cosmological constant.

In this paper we start from sigma model bosonic string action in the presence of the gravitational $G_{\mu\nu}$, Neveu-Schwarz $B_{\mu\nu}$ and dilaton
field $\Phi$ \cite{bozonska}. The bosonic string theory is used here as toy model having in mind that superstring theory corresponds to the real world. We comment quantization of the bosonic string from the viewpoint of the one-loop $\beta$-functions \cite{FCB}. They represent the conditions for background fields under which quantum conformal symmetry is conserved. Conservation of the conformal symmetry on the world-sheet is important because the breaking of symmetry would mean breaking of the two dimensional diffeomorphisms as a symmetry - theory would depend on the choice of the world-sheet coordinates. Demanding quantum conformal invariance and considering one-loop corrections,
we have $\beta^G_{\mu\nu}=\beta^B_{\mu\nu}=0$ and $\beta^\Phi=c$,
where constant $c$ is the central charge of the Virasoro algebra. The equation
$\beta^\Phi=c$ is a consequence of the particular relation between
beta functions (\ref{eq:betarel}).

The quantum conformal invariance can be reached in two ways. The
standard one is putting $c=0$ and $D=26$ and that is a case of critical string \cite{FCB}. In this case
the particular combination of $\beta^G_{\mu\nu}$ and $\beta^\Phi$
represents the equation for $G_{\mu\nu}$ in the form of Einstein
one with some matter. This equation, as well as
$\beta^B_{\mu\nu}=0$ and $\beta^\Phi=0$, can be considered as
equations of motion for the background fields $G_{\mu\nu}$,
$B_{\mu\nu}$ and $\Phi$, respectively. These equations can be
obtained by varying the particular effective string landscape action with respect to
the corresponding background fields \cite{FCB}.

But there is other possibility, $c\neq 0$ and $D\leqslant 26$. The anomaly,
represented by the Schwinger term of the Virasoro algebra, can be
canceled by adding Liouville term to the sigma model action.
Appropriately chosen coefficient in front of Liouville term makes
the theory fully conformally invariant \cite{liuvil}.
The particular combination of $\beta^G_{\mu\nu}$ and
$\beta^\Phi-c$ gives the equation for $G_{\mu\nu}$ in the form of
Einstein one in the presence of some matter and
\textit{cosmological constant}. The central charge of Virasoro algebra takes the role of cosmology constant. This equation with
$\beta^{B}_{\mu\nu}=0$ and $\beta^\Phi=c\;$ can be considered as
equations of motion for background fields $G_{\mu\nu}$,
$B_{\mu\nu}$ and $\Phi$, respectively, because they can be
obtained by varying the string action from \cite{FCB} with
additional, cosmological term. It is obvious that both options, $c=0$ and $c\neq 0$, belong to the string theory landscape \cite{cvafa}, because actions are obtained from consistency conditions.

In general, central charge is an operator which commutes with all other symmetry generators. In the case of twodimensional conformal symmetry that is so called {\it c-number}. Consequently, if we succeed to relate cosmological constant with central charge of the twodimensional conformal symmetry, cosmological constant will be unchanged under world-sheet conformal transformations. But, on the other hand, there are some articles which relate twodimensional conformal symmetry with space-time symmetries, \cite{evov,ljissim}. In these articles one searches for such space-time symmetries under which Virasoro algebra of the energy-momentum tensor components is preserved. Using results of these articles, we can see the possible relation between cosmological constant and space-time symmetries and some interesting consequences of that relation.

At the end we give some closing remarks and comments.

\section{Action and standard conditions for quantum conformal invariance}
\setcounter{equation}{0}

Here we present the known result that standard conditions of
quantum conformal invariance of the bosonic string theory can be
considered as Euler-Lagrange equations of motions obtained from
particular string landscape action \cite{FCB}.

The string action in the presence of gravitational $G_{\mu\nu}$
and $B_{\mu\nu}$ is conformally invariant classically, but on the
quantum level symmetry is broken. In order to keep quantum
conformal invariance the dilaton term is added. The action
describing the string dynamics in the presence of the
gravitational $G_{\mu\nu}$, Neveu-Schwarz $B_{\mu\nu}$ and dilaton
field $\Phi$ is of the form
\begin{equation}\label{eq:action2}
S_{(G+B+\Phi)} = \kappa  \int_\Sigma  d^2 \xi  \sqrt{-g}  \left\{  \left[  {1
\over 2}g^{\alpha\beta}G_{\mu\nu}(x) +{\varepsilon^{\alpha\beta}
\over \sqrt{-g}}  B_{\mu\nu}(x) \right] \partial_\alpha x^\mu
\partial_\beta x^\nu +  \Phi (x) R^{(2)}  \right\} \,  ,
\end{equation}
where $x^\mu$ ($\mu=0,1,\dots,D-1$) are space-time coordinates,
while world-sheet is spanned by
$\xi^\alpha=(\xi^0=\tau,\xi^1=\sigma)$. The symbol $R^{(2)}$
denotes scalar curvature with respect to the world-sheet intrinsic
metric $g_{\alpha\beta}$.

There are three one-loop $\beta$-functions corresponding to the
space-time metric $G_{\mu\nu}$, antisymmetric field $B_{\mu\nu}$,
and dilaton field $\Phi$ \cite{FCB}
\begin{equation}\label{eq:betaG}
\beta^G_{\mu \nu} \equiv  R_{\mu \nu} - \fr{1}{4} B_{\mu \rho
\sigma}
B_{\nu}{}^{\rho \sigma} +2 D_\mu a_\nu    \,  ,
\end{equation}
\begin{equation}\label{eq:betaB}
\beta^B_{\mu \nu} \equiv  D_\rho B^\rho{}_{\mu \nu} -2 a_\rho
 B^\rho{}_{\mu \nu}    \,  ,
\end{equation}
\begin{equation}\label{eq:betaFi}
\beta^\Phi \equiv 2 \pi \kappa {D-26 \over 6} -
 R + \fr{1}{12} B_{\mu \rho \sigma}
B^{\mu \rho \sigma} - 4 D_\mu a^\mu + 4 a^2   \,  ,
\end{equation}
which characterize the conformal anomaly of the sigma model
(\ref{eq:action2}). The space-time Ricci tensor is denoted with
$R_{\mu \nu}$, $B_{\mu \rho \sigma}=\partial_\mu
B_{\nu\rho}+\partial_\nu B_{\rho\mu}+\partial_\rho B_{\mu\nu}$ is
the field strength for the field $B_{\mu\nu}$ and
$a_\mu=\partial_\mu \Phi$ is the gradient of the dilaton field.

Vanishing of these beta functions is the standard condition for
quantum conformal invariance. From these three conditions
equations of motion for background fields $G_{\mu\nu}$,
$B_{\mu\nu}$ and $\Phi$ in $D=26$ follow
\begin{equation}
\left[ B_{\mu\nu}\right]: \beta^B_{\mu\nu}=0\, ,\quad  \left[ \Phi
\right]: \beta^\Phi=0\, ,
\end{equation}
\begin{equation}
\left[ G_{\m\nu}\right]:
\beta^G_{\mu\nu}+\frac{1}{2}G_{\mu\nu}\beta^\Phi=0 \longrightarrow
R_{\mu\nu}-\frac{1}{2}G_{\mu\nu}R=T_{\mu\nu}^{matter}\, ,
\end{equation}
where
\begin{equation}
T_{\mu\nu}^{matter}=\frac{1}{4}B_{\mu\rho\sigma}B_\nu{}^{\rho\sigma}-\frac{1}{24}B^2
G_{\mu\nu}-2D_\mu a_\nu+2G_{\mu\nu}D_\rho a^\rho  - 2 a^2
G_{\mu\nu}\, .
\end{equation}

These equations of motion can be obtained, using variation
principle, from action
\begin{equation}\label{eq:pozadina}
S=\frac{1}{2k_0^2}\int d^Dx \sqrt{-G}e^{-2\Phi}\left[ R+4G^{\mu\nu}a_\mu a_\nu-\frac{1}{12} B_{\mu \rho \sigma}
B^{\mu \rho \sigma}\right]\, ,
\end{equation}
which we will call the effective action.
The normalization constant $k_0$ is not determined by the field
equations and has no physical significance since it can be changed
by a redefinition of $\Phi$.

If we make a conformal scaling of the metric $G_{\mu\nu}(x)$
\cite{FCB1}
\begin{equation}\label{eq:pozadinac1}
\tilde G_{\mu\nu}(x)=e^{2\omega(x)}G_{\mu\nu}(x)\, ,\quad
\omega(x)=2\frac{\Phi_0-\Phi(x)}{D-2}\, ,
\end{equation}
where $\Phi_0$ is a constant, the scalar curvature constructed
from $\tilde G_{\mu\nu}$ is of the form
\begin{equation}
\tilde R=e^{-2\omega}\left[ R-2(D-1)G^{\mu\nu}D_\mu \partial_\nu
\omega-(D-2)(D-1)\partial_\mu \omega \partial^\mu \omega\right]\,
.
\end{equation}

From (\ref{eq:pozadinac1}) we have that $\sqrt{-G}=e^{-\omega
D}\sqrt{-\tilde G}$. Then the action (\ref{eq:pozadina}) takes the
form
\begin{equation}
S_c=\frac{1}{2k^2}\int d^D x \sqrt{-\tilde G}\left[ \tilde
R-\frac{4}{D-2}\tilde G^{\mu\nu}a_\mu a_\nu
-\frac{1}{12}e^{-\frac{8\tilde \Phi}{D-2}}B_{\mu\nu\rho}\tilde
B^{\mu\nu\rho}\right]\, ,
\end{equation}
where $\tilde \Phi=\Phi-\Phi_0$, $\tilde B^{\mu\nu\rho}=\tilde
G^{\mu \sigma}\tilde G^{\nu \epsilon}\tilde G^{\rho
\eta}B_{\sigma\epsilon\eta}$, and $k=k_0 e^{\Phi_0}$. As we can
see the first term in this action is Einstein-like. Because of
that metric $G_{\mu\nu}$ is called sigma model metric, while
$\tilde G_{\mu\nu}$ is Einstein metric.

\section{Adding of Liouville term}
\setcounter{equation}{0}

There is another possibility to establish conformal invariance on
the quantum level. In this section we will present this formalism as an alternative approach to preserve quantum conformal symmetry, and then will find effective theory for the corresponding sigma model action.

In the previous section we saw that sigma model action of bosonic string (\ref{eq:action2}) kept quantum conformal invariance under set of conditions originating from vanishing of beta functions, (\ref{eq:betaG})-(\ref{eq:betaFi}). But, these beta functions are not independent ones and they obey the following identity
\begin{equation}\label{eq:betarel}
D^\mu \beta^G_{\mu\nu}+(4\pi)^2 \kappa D_\nu \beta^\Phi=0\, .
\end{equation}
Consequently, when we choose that $\beta^G_{\mu\nu}=\beta^B_{\mu\nu}=0$, then we have that beta function related to the dilaton field is nonzero constant
\begin{equation}
\beta^\Phi=2\pi \kappa{D-26 \over 6}+4 a^2\equiv c\, .
\end{equation}
Under the conditions $\beta^G_{\mu}=\beta^B_{\mu\nu}=0$ and $\beta^\Phi=c$, the
non-linear sigma model (\ref{eq:action2}) becomes conformal field
theory. There exists a Virasoro algebra with central charge $c$.

An anomaly remains in the theory and it is represented by the Schwinger term of the Virasoro algebra. It can be canceled by the Wess-Zumino term, which in the case of conformal anomaly is a Liouville action
\begin{equation}\label{eq:kinclan}
S_{L}=  -\frac{\beta^\Phi}{2(4\pi)^2\kappa}\int_{\Sigma} d^2 \xi \sqrt{-g}
R^{(2)}\frac{1}{\Delta}R^{(2)}\, ,\quad
\Delta=g^{\alpha\beta}\nabla_\alpha\partial_\beta\, ,
\end{equation}
where $\nabla_\alpha$ is the covariant derivative with respect to
the intrinsic metric $g_{\alpha\beta}$. Appropriate choice of the
coefficient in front of Liouville action makes the theory fully conformally invariant and the complete action takes
the form
\begin{equation}\label{eq:2deo1}
S=S_{(G+B+\Phi)}+S_L \,  .
\end{equation}

Using the conformal gauge
$g_{\alpha\beta}=e^{2F}\eta_{\alpha\beta}$, we obtain $R^{(2)}=-2\Delta F$ and the action (\ref{eq:2deo1}) takes the form
\begin{equation}\label{eq:2deo2}
S=\kappa \int_{\Sigma}d^2 \xi\bigg [\bigg
(\frac{1}{2}\eta^{\alpha\beta}G_{\mu\nu}+\epsilon^{\alpha\beta} B_{\mu\nu}
\bigg)\partial_{\alpha}x^{\mu}\partial_{\beta}x^{\nu}+2\eta^{\alpha\beta}a_\mu \partial_\alpha x^\mu \partial_\beta F+\frac{2}{\alpha}\eta^{\alpha\beta}
\partial_{\alpha}F \partial_{\beta}F\bigg] \,  ,
\end{equation}
where the constant $\alpha$ is defined as
\begin{equation}\label{eq:alfa}
\frac{1}{\alpha}=\frac{\beta^\Phi}{(4\pi\kappa)^2} \, .
\end{equation}

Redefining the conformal factor $F$
\begin{equation}
{}^\star F=F+\frac{\alpha}{2}a_\mu x^\mu
\end{equation}
we get
\begin{equation}\label{eq:2deo}
S=\kappa \int_{\Sigma}d^2 \xi \bigg [\bigg
(\frac{1}{2}\eta^{\alpha\beta}\;{}^\star
G_{\mu\nu}+\epsilon^{\alpha\beta} B_{\mu\nu}
\bigg)\partial_{\alpha}x^{\mu}\partial_{\beta}x^{\nu}+\frac{2}{\alpha}\eta^{\alpha\beta}
\partial_{\alpha}{}^\star F\partial_{\beta}{}^\star F\bigg] \,  .
\end{equation}
We obtained that conformal part of the action is fully decoupled from the rest, while newly introduced metric ${}^\star G_{\mu\nu}=G_{\mu\nu}-\alpha a_\mu a_\nu$ contains information about dilaton. The rest of action is a standard form of the action without dilaton term.

Let us now consider the space-time effective action in the way presented in the previous section. From the conditions for quantum conformal invariance
\begin{equation}
\beta^G_{\mu\nu}=0\, , \quad \beta^B_{\mu\nu}=0\, ,\quad \beta^\Phi-c=0\, ,
\end{equation}
we can construct linear combination
which can be interpreted as equations of motion for the fields
$G_{\mu\nu}$, $B_{\mu\nu}$ and $\Phi$, respectively,
\begin{equation}
\left[ G_{\mu\nu}\right]:
\beta^G_{\mu\nu}+\frac{1}{2}G_{\mu\nu}(\beta^\Phi-c)=0 \, ,
\end{equation}
\begin{equation}
\left[ B_{\mu\nu} \right]: \beta^B_{\mu\nu}=0\, ,
\end{equation}
\begin{equation}
\left[ \Phi \right]: \beta^\Phi-c=0\, .
\end{equation}
Explicit form of the equations is
\begin{equation}\label{eq:jnaGc}
R_{\mu\nu}-\frac{1}{2}G_{\mu\nu} R= T_{\mu\nu}^{matter}+\frac{\tilde c}{2}G_{\mu\nu}\, ,
\end{equation}
\begin{equation}
D_\rho B^\rho{}_{\mu \nu} -2 a_\rho
 B^\rho{}_{\mu \nu} =0
\end{equation}
\begin{equation}\label{eq:jnaFic}
R - \fr{1}{12} B_{\mu \rho \sigma}
B^{\mu \rho \sigma} + 4 D_\mu a^\mu - 4 a^2+\tilde c=0\, ,
\end{equation}
where $\tilde c =c-2\pi\kappa\frac{D-26}{6}$. The most interesting equation is the first one.  
It can be considered as the Einstein equation of
motion with cosmological constant $\tilde c$ and matter described
with energy-momentum tensor $T_{\mu\nu}^{matter}$. Let us write the space-time action which gives 
these equations of motion 
\begin{equation}\label{eq:pozadinac}
S_c=\frac{1}{2k_0^2}\int d^Dx \sqrt{-G}e^{-2\Phi}\left[ R+4G^{\mu\nu}a_\mu a_\nu-\frac{1}{12} B_{\mu \rho \sigma}
B^{\mu \rho \sigma}+\tilde c\right]\, .
\end{equation}
Repeating the procedure from the previous section and making a
conformal scaling of $G_{\mu\nu}(x)$ (\ref{eq:pozadinac1}), the
action (\ref{eq:pozadinac}) takes the form
\begin{equation}
S_c=\frac{1}{2k^2}\int d^D x \sqrt{-\tilde G}\left[ \tilde R-\frac{4}{D-2}\tilde G^{\mu\nu}a_\mu a_\nu -\frac{1}{12}e^{-\frac{8\tilde \Phi}{D-2}}B_{\mu\nu\rho}\tilde B^{\mu\nu\rho}+\tilde c e^{\frac{4\tilde \Phi}{D-2}}\right]\, .
\end{equation}
The last term in the action can not be considered as cosmological one, because it is not a constant and depends on $x^\mu$ through dilaton field. So, the previous statement, considering equations (\ref{eq:jnaGc})-(\ref{eq:jnaFic}), was not correct. We have to make one more assumption.

Choosing dilaton field to be a constant, $\Phi(x)=\Phi_0$, for arbitrary number of dimensions $D\leqslant 26$, we obtain
\begin{equation}\label{eq:sccc}
S_c=\frac{1}{2k^2}\int d^D x \sqrt{- G}\left[ \tilde R -\frac{1}{12} B_{\mu\nu\rho}B^{\mu\nu\rho}+\tilde c \right]\, .
\end{equation}

Here we can identify the Einstein-Hilbert action with some matter originating from metric $G_{\mu\nu}$ and Kalb-Ramond field and cosmological constant 
\begin{equation}\label{eq:coscon}
\Lambda=-\frac{\tilde c}{2}\, .
\end{equation}
The central charge of the Virasoro algebra in this model takes the role of cosmological constant. 

On the level of the world-sheet action (\ref{eq:2deo}), metric ${}^\star G_{\mu\nu}$ is for constant dilaton equal to the metric $G_{\mu\nu}$, while part of the action describing dynamics of the conformal factor is decoupled from the rest. So, in the following analysis of symmetry aspects we consider bosonic, dilaton-free action - the action containing metric $G_{\mu\nu}$ and Kalb-Ramond field $B_{\mu\nu}$. 

\section{Symmetry aspects of the story}
\setcounter{equation}{0}

In the previous section we obtained that cosmological constant is linearly related to the central charge of the world-sheet conformal symmetry. In general, central charge commutes with all generators of the symmetry group, or, in other words, it is invariant under the action of that group. Consequently, in this case, cosmological constant is invariant under world-sheet conformal group. It is interesting to consider some possible connections of the world-sheet symmetry with the space-time symmetries.

The starting point are the papers \cite{evov} and \cite{ljissim}, where the space-time symmetries of the bosonic string propagating in $G_{\mu\nu}(x)$ and $B_{\mu\nu}(x)$ are considered. The main goal of the authors was to find generators of such symmetries, but under one condition. That condition is preservation of the Virasoro algebra i.e. preservation of the world-sheet conformal symmetry. Here we will present some important steps and use the results of the mentioned papers.

\subsection{Canonical Hamiltonian and Virasoro algebra}

Introducing Liouville term and assuming constant dilaton we succeeded to decouple conformal factor and to obtain theory with metric $G_{\mu\nu}(x)$ and Kalb-Ramond field $B_{\mu\nu}(x)$. The action is of the form
\begin{equation}
S=\int d^2\xi \mathcal{L}=\kappa \int d^2\xi \left(\frac{1}{2}\eta^{\alpha\beta}G_{\mu\nu}+\epsilon^{\alpha\beta}B_{\mu\nu}\right)\partial_\alpha x^\mu \partial_\beta x^\nu\, .
\end{equation}
In order to find canonical Hamiltonian, first we have to find canonically conjugated momenta
\begin{equation}
 \pi_\mu=\frac{\partial{\mathcal L}}{\partial \dot x^\mu}=\kappa (G_{\mu\nu}\dot x^\nu-2B_{\mu\nu}x'^\nu)\, ,
\end{equation}
and the canonical Hamiltonian is of the form
\begin{equation}
H_c=\int d\sigma \mathcal H_c\, ,\quad \mathcal H_c=\dot x^\mu \pi_\mu-\mathcal L\, .
\end{equation}
The basic canonical variables are $x^\mu$, $x'^\mu$ and $\pi_\mu$, but it is useful to introduce chiral currents as their combination
\begin{equation}
j_{\pm \mu}=\pi_\mu+2\kappa \Pi_{\pm \mu\nu}x'^\nu\, . \quad (\Pi_{\pm \mu\nu}=B_{\mu\nu}\pm \frac{1}{2}G_{\mu\nu})
\end{equation}
The canonical Hamiltonian gets the more compact form
\begin{equation}\label{eq:Tpm}
\mathcal H_c=T_--T_+\, ,\quad T_{\pm}=\mp \frac{1}{4\kappa}G^{\mu\nu}j_{\pm \mu}j_{\pm \nu}\, .
\end{equation}
Because the coordinates and canonically conjugated momenta satisfy standard Poisson algebra
\begin{equation}\label{eq:spb}
\{x^\mu(\sigma),\pi_\nu(\bar{\sigma})\}=\delta^\mu_\nu\delta(\sigma-\bar\sigma),
\end{equation}
the algebra of the currents is
\begin{eqnarray}
\{j_{\pm\mu}(\sigma),j_{\pm\nu}(\bar\sigma)\}&=&\pm2\kappa
\Gamma_{\mp\mu,\nu\rho}\,
x^{\prime\rho}(\sigma)\delta(\sigma-\bar\sigma)
\pm2\kappa G_{\mu\nu}(x(\sigma))\delta^\prime(\sigma-\bar\sigma),
\nonumber\\
\{j_{\pm\mu}(\sigma),j_{\mp\nu}(\bar\sigma)\}&=&
\pm2\kappa
\Gamma_{\mp\rho,\mu\nu}\,
x^{\prime\rho}(\sigma)\delta(\sigma-\bar\sigma),\label{eq:algebraj}
\end{eqnarray}
where the generalized connection $\Gamma^\mu_{\pm\nu\rho}$ is defined as
\begin{equation}\label{eq:gen}
\Gamma^\mu_{\pm\nu\rho}=\Gamma^\mu_{\nu\rho}
\pm B^\mu_{\ \nu\rho}\, .
\end{equation}
The Christoffel symbol $\Gamma^\mu_{\nu\rho}$ is given by
\begin{equation}
\Gamma^\mu_{\nu\rho}
=\frac{1}{2}(G^{-1})^{\mu\sigma}(\partial_\nu G_{\rho\sigma}+\partial_\rho G_{\sigma\nu}
-\partial_\sigma G_{\nu\rho})\, ,
\end{equation}
while the field strength of the field $B_{\mu \nu}$ is defined as
\begin{equation}
B^\mu_{\ \nu\rho}=(G^{-1})^{\mu\sigma}B_{\sigma\nu\rho}=(G^{-1})^{\mu\sigma}(\partial_\sigma B_{\nu\rho}+\partial_\nu B_{\rho\sigma}+\partial_\rho B_{\sigma\nu})\, .
\end{equation}
Using the algebra of currents (\ref{eq:algebraj}), we obtain that components of energy-momentum tensor $T_{\pm}$ satisfy Virsoro algebra
\begin{eqnarray}\label{eq:vir}
\{T_\pm(\sigma),T_\pm(\bar\sigma)\}&=&-\Big{[}T_\pm(\sigma)+T_\pm(\bar\sigma)\Big{]}\delta^\prime(\sigma-\bar\sigma),
\nonumber\\
\{T_\pm(\sigma),T_\mp(\bar\sigma)\}&=&0\, .
\end{eqnarray}

\subsection{Symmetry aspects}

Let us pay attention on the articles \cite{evov} and \cite{ljissim} and their main result. They have investigated 
the change of the world-sheet energy-momentum tensor as a consequence of 
the change of the background fields, $G_{\mu\nu}$ and $B_{\mu\nu}$. They demanded that transformed energy-momentum tensor
$T_\pm+\delta T_\pm$ still obeys the classical analogue of the Virasoro algebra (\ref{eq:vir}). Here I will briefly reproduce some of the steps from these considerations.

If we make a transformation of the background fields, the current changes as
\begin{equation}
\delta j_{\pm\mu}=2\kappa\delta\Pi_{\pm\mu\nu}(x)x^{\prime\nu},
\end{equation}
and therefore
\begin{equation}\label{eq:deltat}
\delta T_\pm=\frac{1}{2\kappa}\delta\Pi_{\pm\mu\nu}j_\pm^\mu j_\mp^\nu.
\end{equation}
The last expression is obtained using (\ref{eq:Tpm}) and the relation
\begin{equation}
x'^\mu=\pm\frac{G^{\mu\nu}}{2\kappa}(j_{\pm \nu}-j_{\mp \nu})\, .
\end{equation}
In the Ref.\cite{evov} it is shown that this transformation is canonical one, which means that there exists generator $\Gamma$ and the transformation of $T_{\pm}$ (and any other variable) is defined in terms of the Poisson brackets
\begin{equation}
\delta T_\pm=\{\Gamma, T_\pm\}\, .
\end{equation}
This transformation is equal to the 
change of energy-momentum tensor caused by the 
variation of coordinates $x^\mu\rightarrow x^\mu+\delta x^\mu$. Finding the generator we will find at the same time the form of this coordinate transformation. Effectively, we will make a correlation between transformations of the space-time and transformations of the world-sheet which keeps Virasoro algebra i.e. conformal symmetry.

From the point of view of canonical analysis and according to \cite{evov}, only possible generators are canonical momentum $\pi_\mu$ and $x'^\mu$, or equivalently, the currents $j_{\pm\mu}$. So, the generator is of the following form 
\begin{eqnarray}\label{eq:gdef}
\Gamma=\Gamma_++\Gamma_-,
\quad
\Gamma_\pm=\int d\sigma\, \Lambda^\mu_\pm\big(x(\sigma)\big) j_{\pm\mu}(\sigma).
\end{eqnarray}
After short calculation, it is shown that
\begin{equation}
\delta\Pi_{\pm\mu\nu}=\mp\Big(D_{\mp\nu}\Lambda_{\pm\mu}+
D_{\pm\mu}\Lambda_{\mp\nu}\Big)\, ,
\end{equation}
where
\begin{eqnarray}\label{eq:kovdef}
D_{\pm\mu}\Lambda^\nu&=&\partial_\mu \Lambda^\nu+\Gamma^\nu_{\pm\rho\mu}\Lambda^\rho=
D_\mu \Lambda^\nu\pm B^\nu_{\ \rho\mu}\Lambda^\rho,
\end{eqnarray}
is the generalized covariant derivative, while the parameter is chosen in the form
\begin{equation}\label{eq:defxi}
\Lambda_{\pm\mu}=
\frac{1}{2}\xi_\mu\pm\Lambda_\mu\, .
\end{equation}
Finally, the transformations of background fields are
\begin{eqnarray}\label{eq:st}
\delta G_{\mu\nu}&=&-(D_\mu\xi_\nu+D_\nu\xi_\mu),
\nonumber\\
\delta B_{\mu\nu}&=&D_\mu\Lambda_\nu-D_\nu\Lambda_\mu-B_{\mu\nu}^{\ \ \rho}\xi_\rho.
\end{eqnarray}
The transformation of metric corresponds to the transformation under general coordinate transformation (GCT), $x^\mu\to x^\mu+\xi^\mu(x)$. The transformation of Kalb-Ramond field contains two parts: the first one is formally gauge transformation, while the last term is influence of the general coordinate transformation. 

All this consideration holds for closed string. In the case of open string the boundary conditions have to be considered too. In \cite{ljissim} authors showed that the form of the transformations is unchanged, while both $G_{\mu\nu}$ and $B_{\mu\nu}$ obtain corrections in the form of symmetric and antisymmetric field strengths of some newly introduced gauge field $A_\mu$ living on the string endpoints. For Neumann boundary conditions only Kalb-Ramond field gets correction term and it is of the form, $\mathcal B_{\mu\nu}=B_{\mu\nu}+\partial_\mu A_\nu-\partial_\nu A_\mu$. Consequently, we see within this approach why graviton is related with closed string modes, while gauge interactions are related with open string ones.

The general coordinate transformation occurs as a transformation, both in closed and in open string case, which keeps Virasoro algebra of energy-momentum tensor components. In other words, general coordinate transformation of the space-time metric $G_{\mu\nu}$ is some kind of reflection of the world-sheet conformal symmetry. Additionally, in open string sector, gauge fields have to be introduced in order to keep conformal symmetry on the world-sheet. World-sheet conformal symmetry does not change central charge as well as GCT. 

One interesting hypothesis about cosmological evolution has recently occurred \cite{pen} - conformal cyclic cosmology (CCC), where these conclusions might have some importance and meaning. The basic idea of CCC is that evolution of Universe is periodic and the periods are named aeons. Every aeon starts with its own big bang and end of one aeon is the beginning of the next one. The crucial role in the aeon transition period plays space-time conformal symmetry which is symmetry of equations of motion of matter at the end of aeon (massless particles - ultrarelativistic particles and radiation). Roughly speaking, space-time is squeezed by conformal rescaling of metric and the new aeon starts. If we return to our considerations, the space-time general coordinate transformation encompasses conformal transformation, which means that cosmological constant is not changed during aeon transition period. In this way the same (inflatory) evolution of the next aeon is enabled, and that is expected to be according to CCC.

\section{Concluding remarks}
\setcounter{equation}{0}

In this paper we considered the bosonic string in the presence of metric $G_{\mu\nu}$, Kalb-Ramond field $B_{\mu\nu}$ and dilaton field $\Phi$ in the context of the conformal anomaly cancellation. Using obtained results we showed that it is possible to obtain cosmological constant within string theory landscape and related cosmological constant with Virasoro central charge.

Conformal anomaly is expressed via $\beta$ functions $\beta^G_{\mu\nu}$, $\beta^B_{\mu\nu}$ and $\beta^\Phi$. Putting all three one-loop $\beta$ functions to be equal to zero we obtained conditions which background fields must obey in order to keep conformal symmetry on the quantum level. So, these equations could be considered as equations of motion for the background fields and it is possible to write down the action which is known in the literature as action of the effective theory or, shorter, effective action.

But, there is a relation between $\beta^G_{\mu\nu}$ and $\beta^\Phi$ (\ref{eq:betarel}) which enables us to choose $\beta^\Phi$ to be a non-zero constant. That is in fact a central charge of the Virasoro algebra. In order to keep quantum conformal invariance we added Liouville term to the action to cancel Schwinger one. But if we repeat the procedure from the standard case, introducing Einstein frame, we got an effective action which, in the special case of constant dilaton, turns into Einstein-Hilbert action with some matter and cosmological constant $\Lambda=-\frac{\tilde c}{2}$. The cosmological constant is linear function of the Virasoro algebra central charge $\Lambda=A\cdot c+B$ ($A,B$ are constants). 

On the other hand there are authors \cite{evov,ljissim} who related conformal symmetry on the world-sheet with space-time symmetries. The basic demand is to find such space-time symmetries which will not break Virasoro algebra of the energy-momentum tensor components. So, we can conclude, using the results of these papers that general coordinate transformation combined with gauge transformation for $B_{\mu\nu}$ is some kind of reflection of the twodimensional conformal symmetry. We know that twodimensional conformal symmetry does not change central charge, and consequently, cosmological constant. Consequently, we found the set of the space-time transformations which do not change cosmological constant.

It is very important to emphasize that central charge of Virasoro algebra is a constant only in one-loop calculations. That is a case here, because we use one-loop $\beta$-functions. So, this is a limitation of the considerations presented in this article. Going further in loop expansion, central charge of Virasoro algebra is not any more a constant \cite{redlich}.


\begin{thebibliography}{99}

\bibitem{Land} L. D. Landau and M. Lifshitz, {\it The Classical Theory of Fields}, Pergamon Press Ltd., 1975.
\bibitem{misner} C. W. Misner, K. S. Thorne and J. A. Wheeler, {\it
Gravitation}, W. H. Freeman and company, San Francisco, 1973.
\bibitem{sean} S. M. Carroll, {\it Living Rev. Relativity}, {\bf 4}, (2001), 1.
\bibitem{tipIa} Perlmutter, S., et al., {\it Cosmology from type Ia supernovae}, Bull. Am. Astron. Soc. {\bf 29}, 1351 (1997).

\bibitem{bozonska} B. ~Sazdovi\'c, {\it Eur. Phys. J} {\bf C44} (2005) 599; B. ~Nikoli\'c and B. ~Sazdovi\'c, {\it Phys. Rev.} {\bf D74} (2006)
045024.
\bibitem{FCB} C. G. Callan, D. Friedan, E. J. Martinec and M. J. Perry, {\it Nucl. Phys.}
{\bf B 262} (1985) 593;
T. Banks, D. Nemeschansky and A. Sen, {\it Nucl. Phys.} {\bf B 277} (1986) 67.
\bibitem{liuvil} B. ~Nikoli\'c and B. ~Sazdovi\'c, {\it Phys. Rev.} {\bf D75} (2007) 085011; B. ~Nikoli\'c and B. ~Sazdovi\'c, {\it Adv.Theor.Math.Phys.} {\bf 14} (2010) 1, 1-27.

\bibitem{cvafa} C. Vafa, {\it The String Landscape and the Swampland}, arxiv: hep-th/0509212; T. D. Brennana, F Cartab, and C. Vafa, {\it The String Landscape, the Swampland, and theMissing Corner}, Proceedings of Science, Volume 305 - Theoretical Advanced Study Institute Summer School 2017 "Physics at the Fundamental Frontier" (TASI2017), arXiv:1711.00864.
\bibitem{evov} M. Evans, B. Ovrut, {\it Phys. Rev.} {\bf D39} (1989) 3016; {\it Phys. Rev.} {\bf D41} (1990) 3149.

\bibitem{ljissim} Lj. Davidovi\'c, B. Sazdovi\'c, {\it Eur.Phys.J.} {\bf C78} (2018) 600.


\bibitem{FCB1} M. B. Green, J. H. Scwarz
and E. Witten, \textit{Superstring Theory}, Cambridge University
Press, 1987; J. Polchinski, {\it String theory}, Cambridge
University Press, 1998.





\bibitem{pen} R. Penrose, {\it The basic ideas of conformal cyclic cosmology}, AIP Conf. Proc. 1446, 233 (2012); {\it Cycles of Time - An Extraordinary New View of the Universe}, The Bodley Head, London (2010).
\bibitem{redlich} A. N. Redlich, {\it Phys. Rev.} {\bf D33} (1986) 1094.
\end{thebibliography}
\end{document}